\documentclass[aps,prc,superscriptaddress]{revtex4-2}
\usepackage{color,epsfig}
\usepackage{amssymb}
\usepackage{amsmath}
\usepackage{bm}
\usepackage{natbib}
\usepackage[colorlinks=true,citecolor=blue]{hyperref}
\usepackage{subcaption}

\newcommand{\be}{\begin{eqnarray}}
\newcommand{\ee}{\end{eqnarray}}
\newcommand{\nn}{\nonumber\\}

\begin{document}

\title{Charge sum rules for quark fragmentation functions}

\author{D. Kotlorz}
\email {dorota@theor.jinr.ru}
\affiliation{Bogoliubov Laboratory of Theoretical Physics, Joint Institute
for Nuclear Research, Joliot-Curie 6, Dubna, 141980, Russia}
%ORCID: 0000-0002-4755-2924

\author{O. V. Teryaev}
\email {teryaev@theor.jinr.ru}
\affiliation{Bogoliubov Laboratory of Theoretical Physics, Joint Institute
for Nuclear Research, Joliot-Curie 6, Dubna, 141980, Russia}
\affiliation{Veksler and Baldin Laboratory of High Energy Physics, Joint Institute
for Nuclear Research, Joliot-Curie 6, Dubna, 141980, Russia}
%ORCID: 0000-0001-7002-9093

\date{\today}

\begin{abstract}
Charge sum rules for quark fragmentation functions are studied.
The simultaneous implementation of the conservation of electric and baryon
charges, strangeness and isospin symmetry is achieved when the fragmentation
to both mesons and baryons is considered. The results are compatible to
Gell-Mann--Nishijima formulas and may be the new manifestation of superconformal
symmetry between mesons and baryons.
The numerical estimates are performed and compared with phenomenological
models. The recently suggested violations of sum rules due to Wilson line
contributions are discussed.
\end{abstract}

\maketitle

\section{Introduction}
The basic concept underlying the theoretical analysis of most high energy
interactions in quantum chromodynamics (QCD) is factorization.
As a result of the factorization, the physical quantities (e.g. cross sections)
may be given as convolution of the two separated parts:
the long-distance part that contains information on the structure of the nucleon in
terms of its parton distribution functions (PDFs) and fragmentation functions (FFs),
and the short-distance part which describes the hard interactions of the partons.
The long-distance contributions are universal, i.e., they are the same in any
inelastic reaction, and the short-distance parts depend only on the
large scales related to the large momentum transfer and, therefore, can
be evaluated using perturbative methods of QCD.
PDFs and FFs controlled by the nonperturbative dynamics of QCD and determined from
one process can be used for other processes.

The longitudinal fragmentation functions $D_i^h(z,Q^2)$ $(i=q,\bar{q},g)$ are the final-state
analogs of the PDFs indicating the probability density that an outgoing parton $i$
produces a hadron $h$ with the momentum fraction $z$.
While the PDFs are fairly well known, on the contrary, the flavor-separated quark
and gluon FFs, being relatively new objects, required for a quantitative description
of hard scattering processes involving identified light hadrons in the final-state,
are not so well constrained.
There are different sources to extract FFs from experimental data:
semi-inclusive $e^+\,e^-$ annihilation (SIA), single-inclusive production of
a hadron $h$ at a high transverse momentum $p_T$ in hadron-hadron collisions and
unpolarized semi-inclusive deep inelastic lepton-nucleon scattering (SIDIS).
Only the latter are crucial for a reliable determination of FFs, because only
then one can separate $D_q^h(z,Q^2)$ from $D_{\bar{q}}^h(z,Q^2)$ (from the other
processes only the sum of them can be determined).

In analogy to PDFs, FFs obey the various sum rules reflecting the conservation
laws in QCD. These sum rules, among which the best known example is the momentum
sum rule, provide useful phenomenological constraints for the practical
extraction of the fragmentation functions.
At present there are several sets of FFs that well describe data,
but nevertheless differ quite a lot in different kinematic regions,
particularly at very small $z$, where the NLO perturbative QCD evolution may
lead to the unphysical negative $D_q^h(z,Q^2)$. This make the energy and
other sum rules for FFs a delicate concept which cannot be resolved unless
the $z\rightarrow 0$ behavior of FFs is better understood.

In this work, we study the sum rule for the conserved flavor electric
charge for FFs and demonstrate that it hold implying simultaneously the
conservation of the strangeness and baryon number. Based on this approach,
we obtain the constraints for $D_{uval}^h(z,Q^2)$, where $h$ denotes mesons
$\pi$, $K$ and baryons $p$, $n$, $\Lambda$.
We find these constraints confirmed by recent parametrizations of FFs.

We also discuss the possible cancellations of Wilson line
contributions between quarks and antiquarks for valence fragmentation
functions to resolve the problem of validation of the charge sum rules
coming from the small $z\rightarrow 0$ region.

In the next section, we give some introduction to the fragmentation functions
and their sum rules. Section III contains our main results on the charge sum
rule. In Section IV we compare the results with experimental data.
Finally, we summarize our results in Conclusions.

\section{Fragmentation functions}
Parton fragmentation functions $D_i^h(z,Q^2)$ carry information on the
hadronization process and are related to a nonperturbative aspect of QCD.
The fragmentation process of the hadron $h$ at some hard scale $Q^2$
occurs from a parton $i$ with the probability density $D_i^h(z,Q^2)$,
where $z$ is a fraction of the parton energy carrying by the hadron.

The fragmentation functions, similarly to the PDFs, obey the DGLAP $Q^2$
evolution, and the initial $D_i^h(z,Q^2=Q_0^2)$ can be parametrized in
in different ways, usually in the form
\be
&& D_i^h(z,Q_0^2) = \nn
&& \frac{M_i^h\,z^{\alpha_i}(1-z)^{\beta_i}
\left[1+\gamma_i(1-z)^{\delta_i}\right]}{B[2+\alpha_i,\beta_i+1]+
\gamma_iB[2+\alpha_i,\beta_i+\delta_i+1]}\, ,
\label{eq.1}
\ee
where $M_i^h$,
\be
M_i^h < 1\, ,
\label{eq.3}
\ee
is the energy fraction for the hadron $h$ which is created
from the parton $i$, and $B[a, b]$ denotes the Euler Beta-function.

The momentum sum rule,
\be
\sum\limits_h\int\limits_0^1 dz\, z D_i^h(z,Q^2) = 1\, ,
\label{eq.2}
\ee
is valid separately for each flavor $i$ and involves a sum over all possible
produced hadrons.
The momentum sum rule is true at all scales $Q^2$, what is guaranteed by
DGLAP evolution, and reflects the energy conservation.
This is a rigorous assumption used in most phenomenological extractions of
FFs e.g. \cite{Kretzer:2000yf,Hirai:2007cx,Albino:2008fy,deFlorian:2014xna,
deFlorian:2017lwf,Leader:2015hna}.

However, in some recent fits of the fragmentation functions the momentum sum
rule is not imposed {\it a priori} but rather is used as an {\it a posteriori}
check e.g. \cite{Bertone:2017tyb,Gao:2024nkz,Gao:2024dbv}.
There are two important reasons for such a  procedure. First, the momentum sum
rule, Eq.~(\ref{eq.2}), requires the knowledge of the FFs of all produced
hadrons, and second, it requires integration over $z$ down to $z=0$, while
FFs can be usually determined from the experimental data only to
$0.01<z_{\rm min}<0.2$. Hence, extrapolation of the FFs parametrizations to
the small-$z$ region may lead to their unphysical behavior in this region.
The lack of knowledge of the real small-$z$ behavior of FFs causes the
verification of the sum rules to be problematic. This is not so crucial in
the case of the momentum sum rule due to the suppression of small-$z$
contributions but may become essential in the case of the charge and particle
number sum rules.
Also, the low-$z$ problem with the formulation of the charge and number sum
rules for FFs from the theoretical point of view has been recently raised in
\cite{Collins:2023cuo}.

\section{Charge sum rule}
The charge sum rule for FFs under study expresses charge
conservation:
\be
Q_i = \sum\limits_h Q_h\int\limits_0^1 dz\, D_i^h(z,Q^2)\, ,
\label{eq.4}
\ee
where $Q_i$ is the conserved (in particular,  electric) charge of the parent
quark of flavor $i$ and the sum runs over all produced hadrons of charge $Q_h$.
Unlike the case of
the momentum sum rule, Eq.~(\ref{eq.2}), where the suppression of
small-$z$ contributions occurs, the charge sum rule can be invalidated
due to lack of the experimental data and also an adequate theoretical
interpretation of the fragmentation functions in the range of $0<z<z_{min}$.

For a given process of charged-hadron production the small-$z$ range
depends on $Q$, $z_{min}\sim 1/Q$ for large $Q$, and it is reasonable to
consider instead of the charge sum rule, Eq.~(\ref{eq.4}), its truncated
contribution, which depends on the scale $Q^2$:
\be
Q_i(Q^2) = \sum\limits_h Q_h\int\limits_{z_{min}(Q^2)}^1 dz\, D_i^h(z,Q^2)\, .
\label{eq.5}
\ee
In this way, one can test consistency of data with a theoretical model.

Note also that the region of small $ z \leq \frac{\mu^2 x_B }{ Q^2}$
(where $\mu$ is a scale of an order of typical hadron mass) in SIDIS may
correspond \cite{Teryaev:2002wf} to the breaking of factorization of
independent distribution and fragmentation functions and to the appearance of
(extended) fracture
\cite{Trentadue:1993ka,Grazzini:1997ih,deFlorian:1997wi,Anselmino:2011ss}
functions. The simultaneous consideration of sum rules for fragmentation
and fracture functions may be therefore interesting. 

In much of the data for the charged hadron production, the observed hadrons
are identified as one of the three lightest ones: pions ($\pi^{\pm}$),
kaons ($K^{\pm}$) and protons ($p/\bar{p}$).
Here, in a simple approach to the charge sum rule, we consider these
particles by adding step by step the subsequent components.
Using only isospin SU(2) symmetry for the favored
and unfavored fragmentation functions, and also the charge conjugation
invariance of the strong interactions, we arrive at the generalized
conservation law including charge, strangeness and baryon number.
On this basis, we obtain the constraints for the valence fragmentation
functions of quark $u$, $D_{uval}^h(z,Q^2)$, into mesons $\pi^+$ and $K^+$ and
baryons $p$, $n$ and $\Lambda$, and we compare the results with data.

Let us consider the charge sum rule for $up$ and $down$ quarks.
Henceforth, we skip for simplicity the argument $Q^2$ in functions
$D_q^h(z,Q^2)$. Writing out Eq.~(\ref{eq.4}) explicitly, we have
\be
Q_u &=& \sum\limits_h\int\limits_0^1 dz\,
\left(D_{u}^{h^+}(z)-D_{u}^{h^-}(z)\right) \nn
&=& \sum\limits_h\int\limits_0^1 dz\, D_{uval}^{h^+}(z)
\label{eq.6}
\ee
and
\be
Q_d &=& \sum\limits_h\int\limits_0^1 dz\,
\left (D_{d}^{h^+}(z)-D_{d}^{h^-}(z)\right ) \nn
&=& \sum\limits_h\int\limits_0^1 dz\, D_{dval}^{h^+}(z)\, ,
\label{eq.7}
\ee
respectively, where
\be
D_{qval}^{h^+}(z)=D_{q}^{h^+}(z)-D_{\bar{q}}^{h^+}(z)
\equiv D_{q}^{h^+-h^-}(z)\, .
\label{eq.8}
\ee
In the above formulas, Eqs.~(\ref{eq.6})-(\ref{eq.8}), we have used
charge symmetry of FFs:
\be
D_{q}^{h^-}(z)=D_{\bar{q}}^{h^+}(z)\, .
\label{eq.9}
\ee
The charged hadron production is dominated by charged pions, and this
approximation  was used long ago (see p.16 of \cite{Efremov:1980ub}) to
perform the pioneering check of the charge sum rule and get the $u$ quark charge
from semi-inclusive deep inelastic neutrino data.
Therefore, as a first
step, we consider only the pion contributions to the sums over $h$ in
Eqs.~(\ref{eq.6}) and (\ref{eq.7}):
\be
Q_u = \frac{2}{3} = \!\!\!\sum\limits_{h=\pi^{\pm}}
\!\!Q_h\int\limits_0^1 dz\, D_u^h(z) = \int\limits_0^1 dz\, D_{uval}^{\pi^+}(z)
\label{eq.10}
\ee
and
\be
Q_d = -\frac{1}{3} = \!\!\!\sum\limits_{h=\pi^{\pm}}
\!\!Q_h\int\limits_0^1 dz\, D_d^h(z) = \!\!\int\limits_0^1 dz\, D_{dval}^{\pi^+}(z).
\label{eq.11}
\ee
Assuming the isospin SU(2) symmetry for the favored and unfavored pion
fragmentation functions ($\pi^+ = (u\bar d)$, $\pi^- = (d\bar u)$),
\be
D_u^{\pi^+}(z) = D_{\bar d}^{\pi^+}(z), ~~~~~
D_{\bar u}^{\pi^+}(z) = D_{d}^{\pi^+}(z)\, ,
\label{eq.12}
\ee
we have
\be
D_{dval}^{\pi^+}(z) = -D_{uval}^{\pi^+}(z)\, ,
\label{eq.13}
\ee
and hence, using Eqs.~(\ref{eq.6}) and (\ref{eq.7}), one get the incompatible relations:
\be
Q_u = \frac{2}{3} = \int\limits_0^1 dz\ D_{uval}^{\pi^+}(z)\, ; \\
Q_d = - \frac{1}{3} = - \int\limits_0^1 dz\ D_{uval}^{\pi^+}(z)\, .
\label{eq.13a}
\ee
Still, taking their difference divided by 2 , one get
\be
\frac{Q_u-Q_d}{2} = \frac{1}{2} = \int\limits_0^1 dz\ D_{uval}^{\pi^+}(z)\, ,
\label{eq.13b}
\ee
which is nothing else than the sum rule relating the quark isospin projection to
that of pions. This also shows, that the result obtained in the pioneering paper 
\cite{Efremov:1980ub} should be more close to $1/2$ rather than to $2/3$ which is perfectly compatible 
within $30\%$ accuracy of the data, explicitly mentioned in  \cite{Efremov:1980ub} just after the presented result.  

At the same time, taking the sum (on which we will concentrate in what
follows) instead of the difference one has the obvious contradiction
\be
Q_d = -Q_u.
\label{eq.14}
\ee
or, equivalently,
\be
\frac{1}{3} = 0\, .
\label{eq.15}
\ee

Now, let us check if adding to the charge sum rules for $Q_u$ and $Q_d$ also
the kaon contributions will remove the discrepancy Eq.~(\ref{eq.15}).
Thus, taking into account pions and kaons in the sum over $h$ in
Eqs.~(\ref{eq.6}) and (\ref{eq.7}), we can write
\be
Q_u + Q_d &=& \sum\limits_{h=\pi^{\pm},K^{\pm}} \left (\; \cdots \right ) \nn
&=& \int\limits_0^1 dz\,
\left( D_{uval}^{K^+}(z) + D_{dval}^{K^+}(z)\right)\, ,
\label{eq.16}
\ee
where, as we have shown above in Eq.~(\ref{eq.14}), the pion contribution
to $Q_u+Q_d$ is $0$.
By using the isospin SU(2) symmetry for the kaon fragmentation
functions
%($K^+ = (u\bar s)$, $K^- = (s\bar u)$),
\be
D_{dval}^{K^+}(z) = D_{uval}^{K^0}(z) \equiv D_{u}^{K^0}(z) - D_{u}^{\bar K^0}(z),
\label{eq.17}
\ee
Eq.~(\ref{eq.16}) takes the following form:
\be
Q_u + Q_d = \int\limits_0^1 dz\,
\left(D_{uval}^{K^+}(z) + D_{uval}^{K^0}(z)\right) \, .
\label{eq.18}
\ee
Note that the appearing combination of kaon fragmentation functions should be
zero due to the strangeness (equal zero for light quarks) conservation, so
that the discrepancy persists. Note that the fragmentation functions
of $K^0$ mesons are added here as a matter of principle. They correspond to
the states of different quark content, while experimentally the ones with
definite CP-symmetry are measured, for which the relevant valence contributions
are zero, up to small CP-symmetry violation.

Bearing in mind the above conclusion and also the motivation to find some
general law in terms of the fragmentation functions, we continue our analysis
by including also {\it baryons}.
Thus, after taking into account the contributions to the charge sum rules,
Eqs.~(\ref{eq.6}) and (\ref{eq.7}), coming from pions, kaons and also protons,
we obtain
\be
Q_u + Q_d = \int\limits_0^1 dz\,
\Big( &D&_{u}^{K^+-K^-}(z) +D_{u}^{K^0-\bar K^0}(z) \nn
&+& D_{u}^{p-\bar{p}}(z) +  D_{d}^{p-\bar{p}}(z) \Big)\, .
\label{eq.22}
\ee
Assuming the charge and isospin SU(2) symmetry for the proton ($p=(uud)$)
and neutron ($n=(ddu)$) FFs,
\be
D_d^{p}(z) = D_{u}^{n}(z), ~~~~~
D_{\bar{d}}^{p}(z)\approx D_{\bar{u}}^{p}(z)
\approx D_{\bar{u}}^{n}(z)\, ,
\label{eq.23}
\ee
we can write Eq.~(\ref{eq.22}) as
\be
\frac{1}{3} &=& \langle D_{u}^{K^+}\rangle - \langle D_{u}^{K^-}\rangle
+ \langle D_{u}^{K^0}\rangle - \langle D_{u}^{\bar K^0}\rangle \nn
&+&\langle D_{u}^{p}\rangle - \langle D_{u}^{\bar{p}}\rangle +
\langle D_{u}^{n}\rangle - \langle D_{u}^{\bar{n}}\rangle\, ,
\label{eq.24}
\ee
where we have used the hadron multiplicity in the form
\be
\langle D_{q}^{h}\rangle \equiv \int\limits_0^1 dz\, D_{q}^{h}(z)\, .
\label{eq.25}
\ee

Note that four last terms now express the {\it baryon} charge so that the
current sum rule simultaneously provides conservation of strangeness,
electric and baryon charges, which is a sort of manifestation of the
Gell-Mann--Nishijima formula.

As soon as in the "favored" approximation the fragmentation function of
$K^+$ dominates (and the strangeness conservation is numerically questionable),
it is instructive to add and subtract to the sum rule the same term
(which is also theoretically attracting providing the baryonic strangeness
contribution in addition to the mesonic one)
$\langle D_{u}^{\Lambda}\rangle - \langle D_{u}^{\bar{\Lambda}}\rangle$,
where $\Lambda=(uds)$ hyperon, and then we group all terms in
{\it strange} and {\it non-strange} parts:
\be
\frac{1}{3} &=& \overbrace{\langle D_{u}^{K^+}\rangle - \langle D_{u}^{K^-}\rangle +
\langle D_{u}^{K^0}\rangle - \langle D_{u}^{\bar K^0}\rangle +
\langle D_{u}^{\bar{\Lambda}}\rangle - \langle D_{u}^{\Lambda}\rangle}^
{\text{\large $S_u=0$}} \nn
&+& \underbrace{\langle D_{u}^{p}\rangle - \langle D_{u}^{\bar{p}}\rangle +
\langle D_{u}^{n}\rangle - \langle D_{u}^{\bar{n}}\rangle +
\langle D_{u}^{\Lambda}\rangle - \langle D_{u}^{\bar{\Lambda}}\rangle}_
{\text{\large $B_u=1/3$}}\, .
\label{eq.26}
\ee
One can see that the r.h.s. of Eq.~(\ref{eq.26}) is a sum of the strangeness
$S=0$ and the baryon number $B=1/3$ of the $u$ quark (like in the
Gell-Mann--Nishijima formula), and the final result reads
\be
\frac{1}{3} = \frac{1}{3}\, .
\label{eq.27}
\ee

We have demonstrated that the charge sum rules for the quark fragmentation
functions hold including simultaneously the contributions of mesons and
baryons providing the conservation of the strangeness, electric and baryon
charges.

Let us, for completeness, write the expression for the isospin conservation,
$I_3(u)$, including the pions, kaons and baryons:
\be
\frac{1}{2} &=& \langle D_{u}^{\pi^+}\rangle - \langle D_{u}^{\pi^-}\rangle \nn
            &+& \frac{1}{2} \Big( \langle D_{u}^{K^+}\rangle
-\langle D_{u}^{K^-}\rangle - \langle D_{u}^{K^0}\rangle
+ \langle D_{u}^{\bar K^0}\rangle \nn
            &+& \langle D_{u}^{p}\rangle - \langle D_{u}^{\bar{p}}\rangle
-\langle D_{u}^{n}\rangle + \langle D_{u}^{\bar{n}}\rangle\,\Big)\, .
\label{eq.27a}
\ee
The last two equations form our main result. They are clearly compatible to
Gell-Mann--Nishijima formulas for quarks and hadrons, and it is crucial
that fragmentation to both mesons and baryons must be considered.
This may provide a new manifestation of superconformal symmetry
\cite{Dosch:2015nwa} between mesons and baryons.

Another interesting aspect of the obtained sum rules is provided by the
recent investigation \cite{Collins:2023cuo} of the sum rule violations due to
Wilson line contributions implied by quark and baryon quantum numbers mismatch.
Our result suggests that these effects are strongly constrained by
Gell-Mann--Nishijima formulas.
The simplest way of the realization of such constraints is probably the
cancellations of Wilson line contributions between quarks and antiquarks for
valence fragmentation functions.
This possibility may be supported by the fact \cite{Kotikov:2020ukv}, that
resummed evolution kernels for multiplicities in quark and gluon jets (that is,
the first moments of quark and gluon fragmentation functions) obey the Casimir
scaling. As the gluons' fragmentations to particles and antiparticles coincide,
their quark multiplicities evolutions are the same, due to Casimir scaling.
One may therefore expect that the Wilson line contribution \cite{Collins:2023cuo} is
also the same, although additional investigations are required. It is
interesting to mention, that Casimir scaling in \cite{Kotikov:2020ukv} is
related to supersymmetric properties of evolution, which may be compared with
superconformal symmetry in \cite{Dosch:2015nwa}.

%The crucial point of our approach is the use of FFs as the distribution
%functions to express the strangeness and baryon number of the quark:
%\be
%S_q &=& \sum\limits_{h} S_h\, \langle D_{q}^{h}\rangle \, , \\
%B_q &=& \sum\limits_{h} B_h\, \langle D_{q}^{h}\rangle \, .
%\label{eq.28}
%\ee

%From Eq.~(\ref{eq.26}) one immediately obtains the Gell-Mann--Nishijima
%relation for $u+d$:
%\be
%Q_u + Q_d = \frac{1}{2}\, \left ( S_u + S_d + B_u + B_d \right )\, .
%^\label{eq.30}
%\ee

\section{Comparison to data}
To test consistency of our results with a phenomenological model of the
fragmentation functions we compare the theoretical predictions
on the charge and isospin sum rules with the numerical estimates of
their truncated at $z$ contributions based on some recent parametrizations
of FFs.

In addition to our main results on the charge and isospin sum rules,
below we obtain some constraints on $\langle D_{uval}^{h}\rangle$,
where $h=\pi^+, K^+, p\;{\rm and}\;\Lambda$, which can also be tested
numerically. Namely, gathering together Eqs.~(\ref{eq.24}), (\ref{eq.26})
and also the expression for the electric charge $Q_u$:
\be
\langle D_{uval}^{K^+}\rangle + \langle D_{uval}^{p}\rangle +
\langle D_{uval}^{n}\rangle = \frac{1}{3}\, , \\
\langle D_{uval}^{p}\rangle + \langle D_{uval}^{n}\rangle +
\langle D_{uval}^{\Lambda}\rangle = \frac{1}{3}\, , \\
\langle D_{uval}^{\pi^+}\rangle + \langle D_{uval}^{K^+}\rangle +
\langle D_{uval}^{p}\rangle = \frac{2}{3}\, ,
\label{eq.31}
\ee
and assuming (based on the concept of a common function for
favored fragmentation functions from up and down quarks and on a flavor
symmetry)
\be
D_{uval}^{p}(z)\approx 2D_{uval}^{n}(z)
\approx 2D_{uval}^{\Lambda}(z)\, ,
\label{eq.32}
\ee
we arrive at
\be
\langle D_{uval}^{\pi^+}\rangle = \frac{5}{12}\, , ~~~~~~~
\langle D_{uval}^{p}\rangle = \frac{1}{6}\, , \nn
\langle D_{uval}^{K^+}\rangle = \langle D_{uval}^{n}\rangle =
\langle D_{uval}^{\Lambda}\rangle = \frac{1}{12}\, .
\label{eq.33}
\ee

In our analysis we use the following various published NLO parametrization
sets:
HKNS-2007~\cite{Hirai:2007cx}, AKK-2008~\cite{Albino:2008fy},
DSEHS-2014~\cite{deFlorian:2014xna} and LSS-2015~\cite{Leader:2015hna}
for pions; HKNS-2007, AKK-2008 and DEHSS-2017~\cite{deFlorian:2017lwf}
for kaons; HKNS-2007, DSS-2007~\cite{deFlorian:2007ekg}, AKK-2008 and
BS-2003~\cite{Bourrely:2003wi} for protons; and
DSV-1998~\cite{deFlorian:1997zj}, BS-2003, AKK-2008 and
SAK-2020~\cite{Soleymaninia:2020ahn} for $\Lambda$.
For the charged hadrons, we also employ a very recent set of FFs,
NPC23~\cite{Gao:2024dbv,NPC23}.
A multitude of processes have been used for the extraction of FFs.
The fits are based on SIA (all sets but LSS), SIDIS (LSS, DSEHS,
DEHSS, DSS, NPC23) and $pp(\bar{p})$ (AKK, BS, DSEHS, DEHSS, DSS, NPC23).
DGLAP NLO evolution of FFs has been performed with the help of
a correspondingly modified numerical code provided by HKNS \cite{Hirai:2007cx}.

\begin{figure*}[t]
\centering
\includegraphics[width=0.475\textwidth]{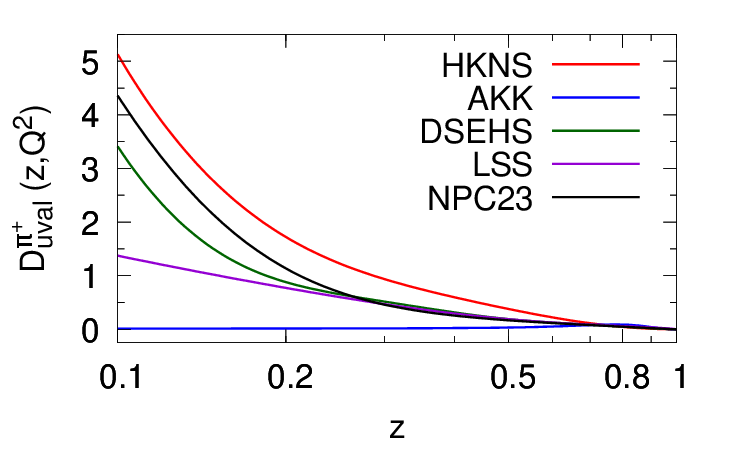}
\hfill
\includegraphics[width=0.475\textwidth]{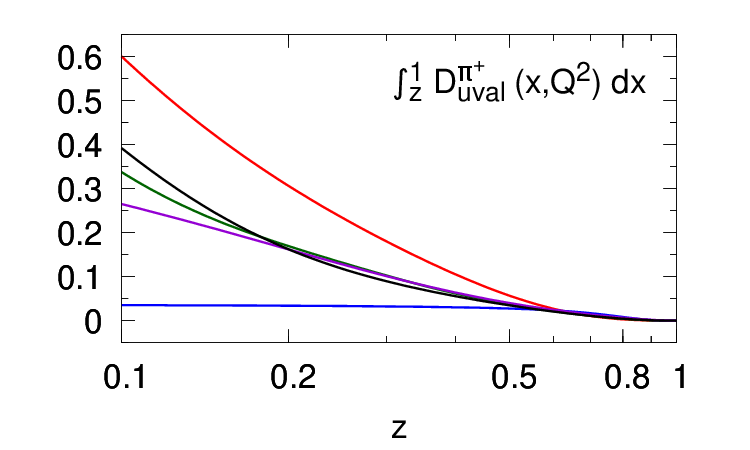}
\caption{The valence fragmentation function $D_{uval}^{\pi^+}(z,Q^2)$ (left)
and its contribution to the charge of the $u$ quark,
$\int_z^1 D_{uval}^{\pi^+}(x,Q^2)\, dx$ (right), for different
parametrization sets at $Q^2=25\,{\rm GeV^2}$.}
\label{nfig1}
\end{figure*}

A comparison of the valence $u$ quark FF for the pion at the scale
$Q^2=25\,{\rm GeV^2}$, $D_{uval}^{\pi^+}(z,Q^2)$, for different
parametrization sets is shown in Fig.~\ref{nfig1} (left).
The initial scales $Q_0^2$ are $1\,{\rm GeV^2}$ for HKNS, LSS and DSEHS,
$2\,{\rm GeV^2}$ for AKK and $25\,{\rm GeV^2}$ for NPC23 fits, respectively.
The AKK FF was extracted directly in the valence ($u-\bar{u}$) form from
measurements in which particles were distinguished from their antiparticles
and has a high uncertainty \cite{Albino:2008fy}.
Large differences in the small-$z$ behavior of $D_{uval}^{\pi^+}$ between
various fits reflect in the predictions for its truncated contributions to
the charge sum rule, $\int_z^1 D_{uval}^{\pi^+}(x,Q_0^2)\, dx$, see
Fig.~\ref{nfig1} (right).
% Figure 1 ****************************************
% Figure 2 ****************************************
\begin{figure*}
        \centering
        \begin{subfigure}[b]{0.475\textwidth}
            \centering
            \includegraphics[width=1\textwidth]{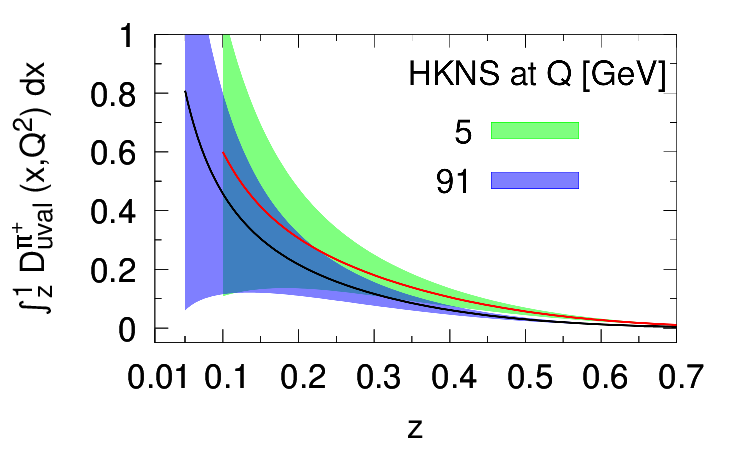}
        \end{subfigure}
        \hfill
        \begin{subfigure}[b]{0.475\textwidth}  
            \centering 
            \includegraphics[width=1\textwidth]{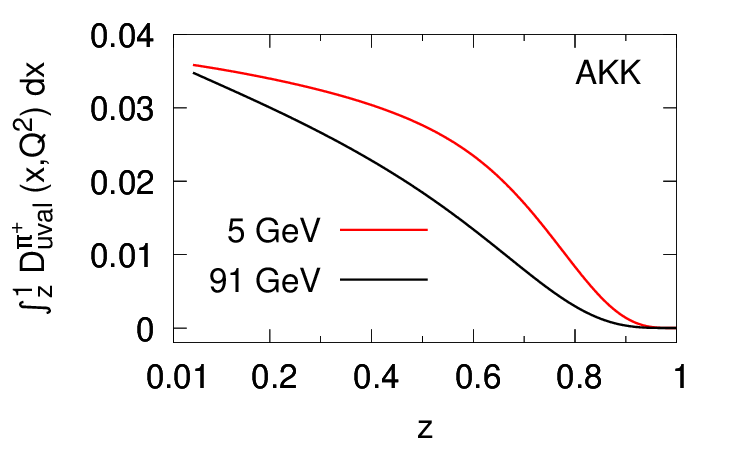}
        \end{subfigure}
        \vskip\baselineskip
        \begin{subfigure}[b]{0.475\textwidth}   
            \centering 
            \includegraphics[width=1\textwidth]{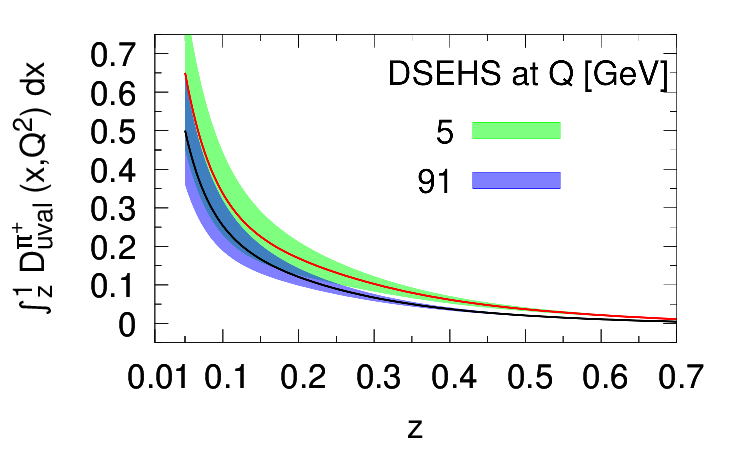}
        \end{subfigure}
        \hfill
        \begin{subfigure}[b]{0.475\textwidth}   
            \centering 
            \includegraphics[width=1\textwidth]{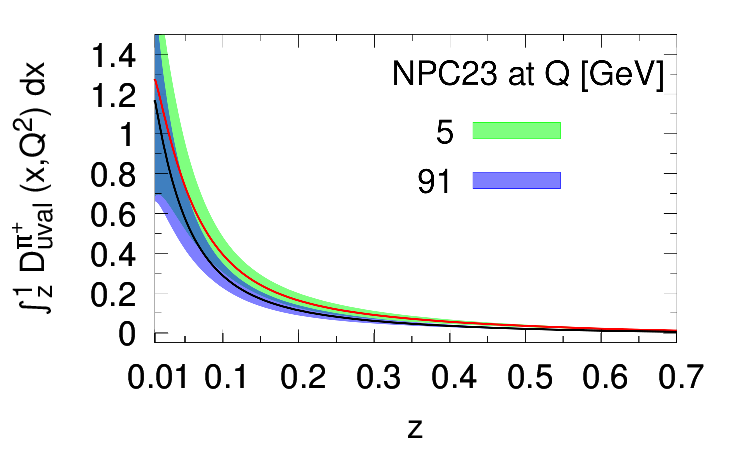}
        \end{subfigure}
\caption{$Q^2$ dependence of the truncated moment
$\int_z^1 D_{uval}^{\pi^+}(x,Q^2)\, dx$ for HKNS, AKK, DSEHS and NPC23 fits.
The range of $z$ is limited to $z_{min}<z<1$ for each fit, respectively.
The uncertainty bands are shown.}
\label{nfig2}
\end{figure*}

Only in the LSS fit the conservation of the electric charge has been imposed
{\it a priori} with the assumption $\langle D_{uval}^{\pi^+}\rangle = 2/3$.
However the increase of the other FFs at small values of $z$ is
phenomenologically not relevant as their range of applicability is restricted
by kinematics in SIA, SIDIS and hadron-hadron collisions to $z>z_{min}$.
Thus, for the HKNS FFs, $z_{min}=0.05$ for the data at $Q=M_Z$, and
$z_{min}=0.1$ for the data at $Q<M_Z$. In the case of the AKK and DSEHS
parametrizations, $z_{min}=0.05$, while for the NPC23 fit $z_{min}=0.01$
for pion and $z_{min}= 0.088$ for kaon and proton FFs. This is taken into
account in the following figures illustrating our results to avoid dealing with
an unphysical extrapolation of FFs to the region $z<z_{min}$.

Figure~\ref{nfig2} shows the truncated pion contribution to the charge sum
rule, $\int_z^1 D_{uval}^{\pi^+}(x,Q^2)\, dx$ for HKNS, AKK, DSEHS and NPC23
fits at $Q=5\,{\rm GeV}$ and $Q=M_Z$ together with the uncertainty bands,
for the restricted range of $z_{min}<z<1$, respectively.
The charge sum rule, Eq.~(\ref{eq.4}), which reflects the electric charge
conservation holds at any scale $Q^2$ while its truncated at $z_{min}$ version,
Eq.~(\ref{eq.5}), depends on $Q^2$. The HKNS results suffer from significantly
larger uncertainties, estimated by the Hessian method, at small-$z$ in
comparison to DSEHS and NPC23 fits.
In turn, the AKK fit which at the initial scale $Q_0^2=2\,{\rm GeV^2}$ has
a simple form, $D_{uval}^{\pi^+}(z,Q_0^2)\sim z^{13}(1-z)^{2.5}$, definitely
underestimates the pion contribution to the charge sum rule.

\begin{figure*}
        \centering
        \begin{subfigure}[b]{0.475\textwidth}
            \centering
            \includegraphics[width=1\textwidth]{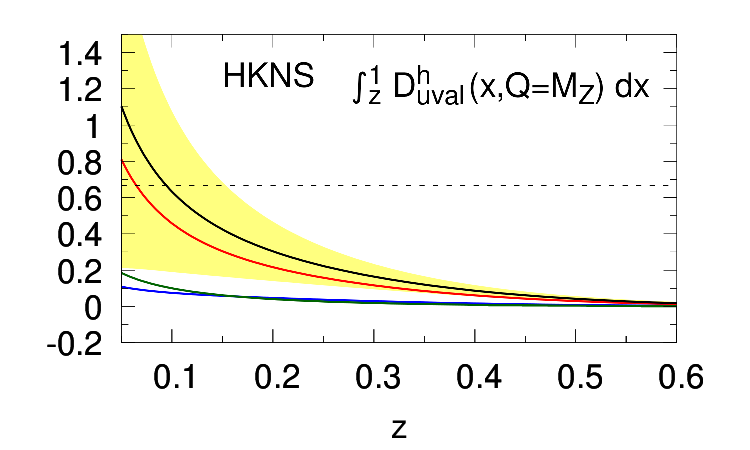}
        \end{subfigure}
        \hfill
        \begin{subfigure}[b]{0.475\textwidth}  
            \centering 
            \includegraphics[width=1\textwidth]{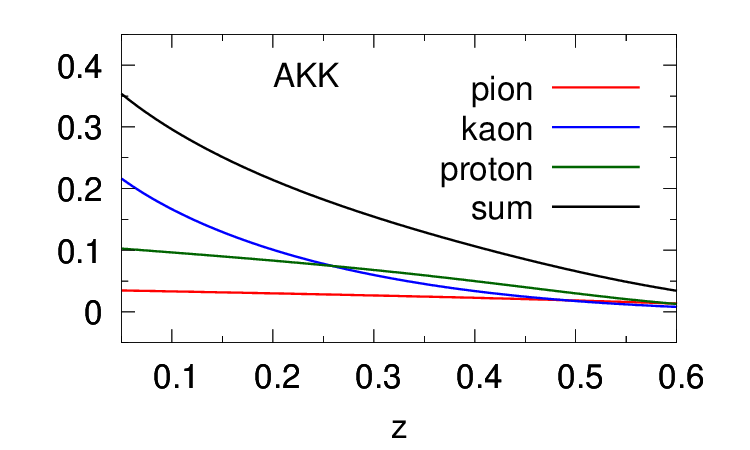}
        \end{subfigure}
        \vskip\baselineskip
        \begin{subfigure}[b]{0.475\textwidth}   
            \centering 
            \includegraphics[width=1\textwidth]{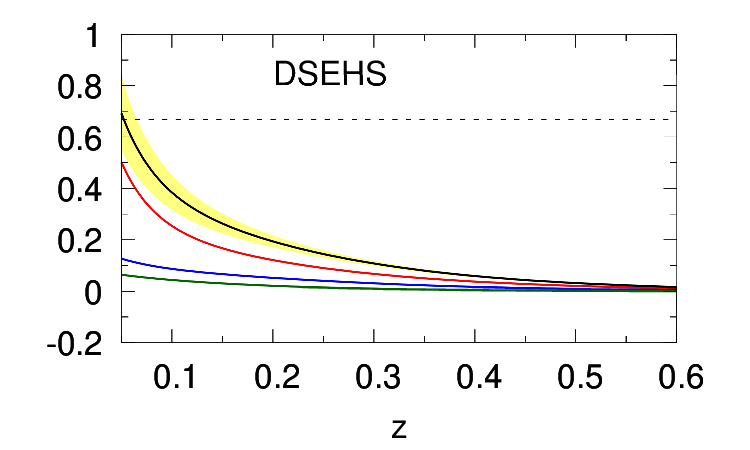}
        \end{subfigure}
        \hfill
        \begin{subfigure}[b]{0.475\textwidth}   
            \centering 
            \includegraphics[width=1\textwidth]{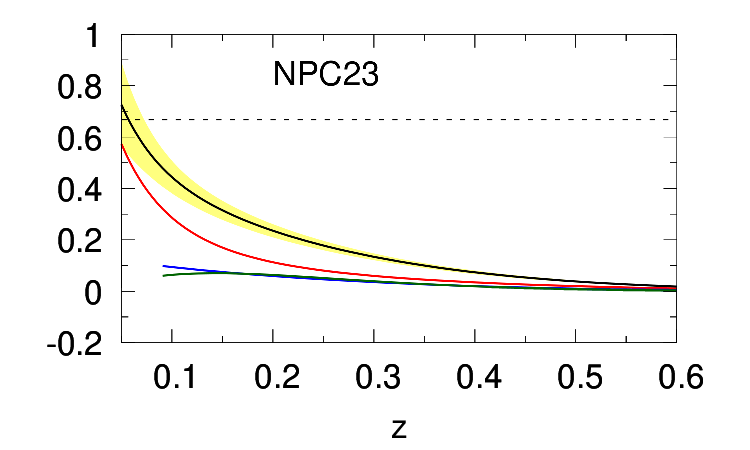}
        \end{subfigure}
\caption{Contributions to the charge of the $u$ quark coming from the pions,
kaons and protons, $\int_z^1 D_{uval}^{h}(x,Q^2)\, dx$, and their sum,
Eq.~(\ref{eq.31}), for different FFs fits at $Q=M_Z$. The uncertainty band
for the sum is shown.}
\label{nfig3}
\end{figure*}

The separate contributions of the pion, kaon, proton and their sum to the
electric charge of the $u$ quark ($Q_u=2/3$) is shown in Fig.~\ref{nfig3}.
One can see that the predictions of the HKNS, NPC23 and, particularly,
DSEHS (DEHSS, DSS) fits are in good agreement with the theoretical charge
sum rule, Eq.~(\ref{eq.4}).
% Figure 3 ****************************************
% Figure 4 ****************************************
\begin{figure*}
\centering
\includegraphics[width=0.475\textwidth]{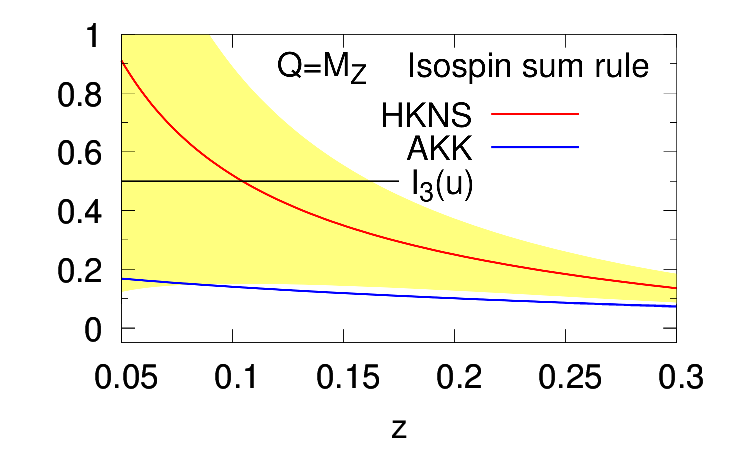}
\hfill
\includegraphics[width=0.475\textwidth]{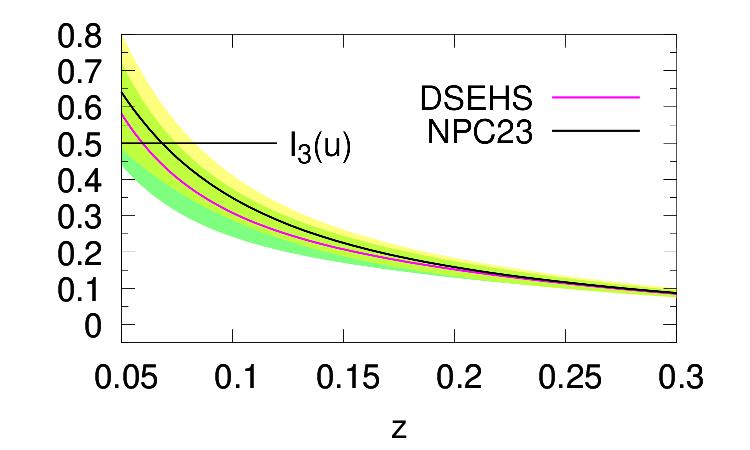}
\caption{Truncated contributions to the isospin sum rule for the $u$ quark,
Eq.~(\ref{eq.34}).}
\label{nfig4}
\end{figure*}

In Fig.~\ref{nfig4} we present the truncated at $z_{min}$
contributions to the isospin sum rule for the $u$ quark utilizing the HKNS, AKK,
DSEHS (DEHSS, DSS) and NPC23 FFs. Based on Eq.~(\ref{eq.27a}), which
incorporates both the mesons and baryons in the sum for $I_3(u)$, and using
the relations Eqs.~(\ref{eq.17}) and (\ref{eq.32}), where it is usually assumed
that $D_{dval}^{K^+}=0$, one obtains
\be
I_3(u)\approx\langle D_{uval}^{\pi^+}\rangle +
\frac{1}{2}\langle D_{uval}^{K^+}\rangle +
\frac{1}{4}\langle D_{uval}^{p}\rangle \, .
\label{eq.34}
\ee

The value $I_3(u)=0.5$ is reached at $z\approx 0.1$ for the HKNS result and
near $z_{min}$ for the DSEHS and NPC23 predictions. The AKK fit of $I_3(u)$,
similarly as in the case of the electric charge sum rule, lies significantly
below the theoretical value.\\
Concluding, the data on the final-state hadronization encoded in the tested
parton-to-hadron FFs are mostly in agreement with the theoretical charge
and isospin sum rules.

\begin{figure*}
        \centering
        \begin{subfigure}[b]{0.475\textwidth}
            \centering
            \includegraphics[width=1\textwidth]{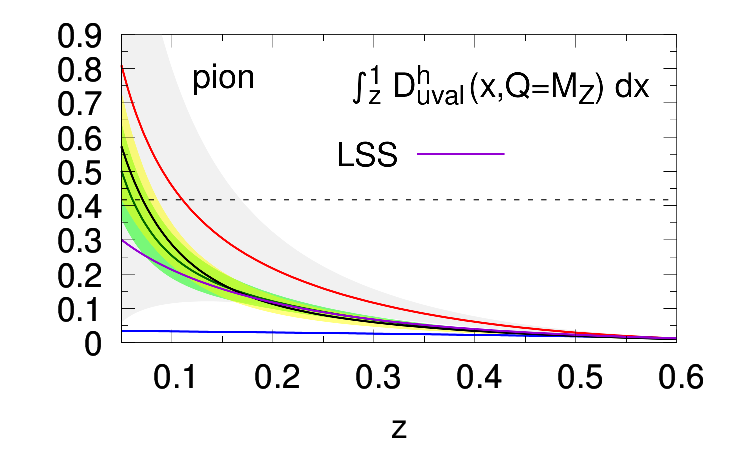}
        \end{subfigure}
        \hfill
        \begin{subfigure}[b]{0.475\textwidth}  
            \centering 
            \includegraphics[width=1\textwidth]{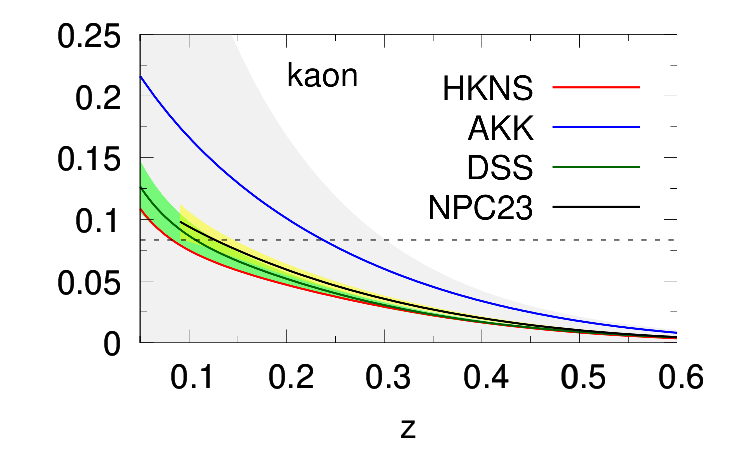}
        \end{subfigure}
        \vskip\baselineskip
        \begin{subfigure}[b]{0.475\textwidth}   
            \centering 
            \includegraphics[width=1\textwidth]{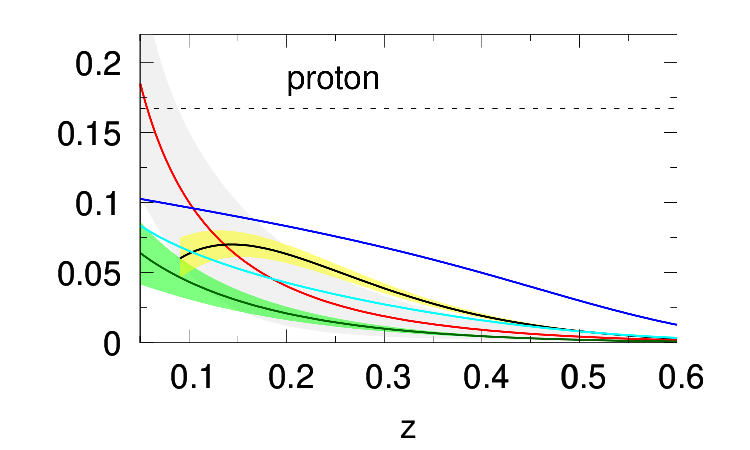}
        \end{subfigure}
        \hfill
        \begin{subfigure}[b]{0.475\textwidth}   
            \centering 
            \includegraphics[width=1\textwidth]{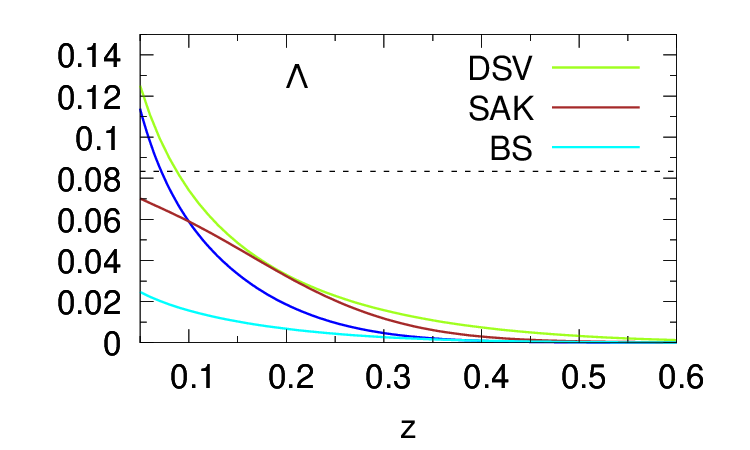}
        \end{subfigure}
\caption{The truncated moments, $\int_z^1 D_{uval}^{h}(x,Q^2)\, dx$,
for mesons (the pions and kaons) and baryons (the protons and $\Lambda$),
calculated for different FFs sets, compared to
the constraints for $\langle D_{uval}^{h}\rangle$, Eq.~(\ref{eq.33}).
The uncertainty bands are shown.}
\label{nfig5}
\end{figure*}

It is also interesting to notice that the constraints on
$\langle D_{uval}^{h}\rangle$ obtained in Eq.~(\ref{eq.33}) are supported
by the phenomenological estimates as well (within systematic uncertainties),
with the exception of AKK parametrizations for the pion and kaon.
This is shown in Fig.~\ref{nfig5}, where we present the truncated at $z$
first moment of $D_{uval}^h$ for mesons: pions and kaons; and for baryons:
protons and $\Lambda$. However, the comparison for the $\Lambda$ should be treated
with caution as the data and hence the obtained fits are accessible only for
the sum $\Lambda + \bar{\Lambda}$. Therefore, our assumption that
$D_{\bar{u}}^{\Lambda}\ll D_{u}^{\Lambda}$, and also the assumption from
Eq.~(\ref{eq.32}) which relates $D_{uval}^{\Lambda}$ to $D_{uval}^{p}$
may be too rough.

% Figure 5 ****************************************
% Figure 6 ****************************************
\begin{figure*}
\centering
\includegraphics[width=0.475\textwidth]{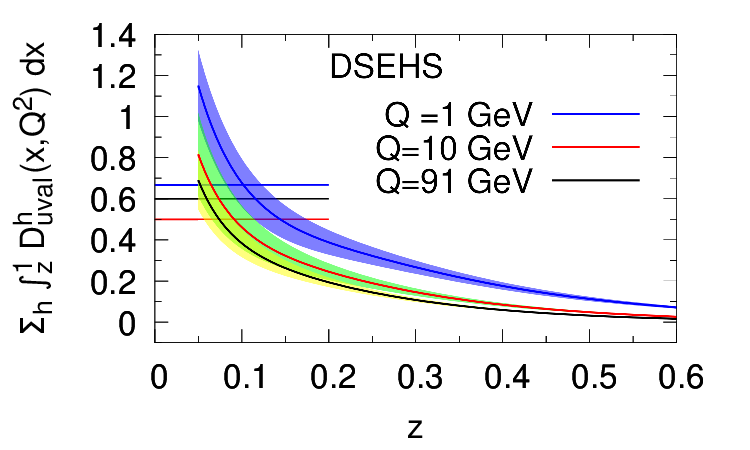}
\hfill
\includegraphics[width=0.475\textwidth]{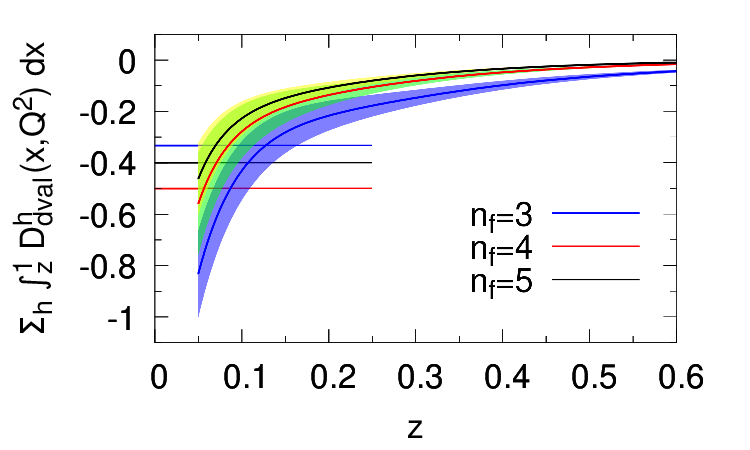}
\caption{Truncated contributions to the electric charge of the $u$ quarks
and the $d$ quarks for DSEHS (DEHSS, DSS) FFs at different scales $Q^2$.
Comparison to the modified charge sum rule, Eq.~(\ref{eq.35}),
\cite{Collins:2023cuo} (horizontal lines). Different $n_f$ numbers of active
flavors of the so-called "orphan quarks" at $z=0$ correspond to the scales
$Q^2$, respectively.}
\label{nfig6}
\end{figure*}

In our analysis we avoid dealing with the region of the small
$z\rightarrow 0$, where the parton to hadron fragmentation functions are
unknown, by using the truncated at $z_{min}\sim 0.05$ contributions to the
charge and isospin sum rules. We note a good agreement of the
phenomenological results with the theoretical predictions near the smallest
accessible experimental $z_{min}$.\\
Our results seem to favor the possible cancellations of Wilson line
contributions between quarks and antiquarks for valence fragmentation
functions. This resolves the problem of violation of the charge sum rules
for FFs coming from the final-state hadronization at $z\rightarrow 0$
raised in \cite{Collins:2023cuo}.
Nevertheless, let us finally present a possible modification of the charge
sum rule proposed in \cite{Collins:2023cuo} and its impact on our tests.

To make the charge sum rule valid despite the mismatch of final state quark
numbers, its modified form incorporates the so-called "orphan quarks"
$j$ at $z=0$:
\be
\sum\limits_h Q_h\int\limits_0^1 dz\, D_i^h(z,Q^2) \approx
Q_i-\frac{1}{n_f}\,\sum\limits_{j\,{\rm active}} Q_j\, ,
\label{eq.35}
\ee 
where $n_f$ is the number of all active $j$-quark flavors.
In this way, the modified charge sum rule, Eq.~(\ref{eq.35}), in a contrary
to the standard one, Eq.~(\ref{eq.4}), depends on the scale $Q^2$.
This is shown in Fig.~\ref{nfig6} where we compare the truncated contributions
to the charge sum rules for the $u$ and $d$ quarks for DSEHS (DEHSS, DSS) FFs
at different scales $Q^2$ with the modified charge sum rule, Eq.~(\ref{eq.35}),
for different numbers of active flavors $n_f$ of the so-called "orphan quarks"
at $z=0$.
The scales $Q=1\,{\rm GeV}$, $10\,{\rm GeV}$ and $91\,{\rm GeV}$ are taken
to be below the charm threshold, $Q^2<m_c^2$, between the charm and bottom
ones, $m_c^2<Q^2<m_b^2$, and above the bottom one, $Q^2>m_b^2$,
respectively. These correspond to $n_f=3$, $4$ and $5$, respectively.
If the only active quark flavors are $u$, $d$ and $s$ ($n_f=3$) then the
subtracted term in Eq.~(\ref{eq.35}) vanishes and the charge sum rule,
Eq.~(\ref{eq.4}) is recovered. If $n_f=4$ or $5$, one has
$Q_i\approx 0.5$ or $0.6$ for the $u$ quark, and $Q_i\approx -0.5$ or $-0.4$
for the $d$ quark.

The valence FFs under study, being nonsinglets, do not mix with other FFs
in their $Q^2$ evolution and this property holds in all orders in pQCD.
Therefore the thresholds for heavy quarks enter into evolution only via
the running coupling constant $\alpha_s(Q^2)$.
From Fig.~\ref{nfig6} we observe that the modification of the charge rule,
Eq.~(\ref{eq.35}), does not have much impact on the results of our
phenomenological analysis.

\section*{Conclusions}
We demonstrated that the charge sum rules for the quark fragmentation
functions hold including simultaneously the contributions of mesons and
baryons providing the conservation of the strangeness, electric and baryon
charges, Eq.~(\ref{eq.26}). We also obtained the expression for the isospin
conservation, Eq.~(\ref{eq.27a}).
The results are compatible to Gell-Mann--Nishijima formulas for quarks and
hadrons manifesting a new aspect of quark-hadron duality.

Using our results, we formulated the constraints for $D_{uval}^h(z,Q^2)$,
Eq.~(\ref{eq.33}), where $h$ denotes mesons $\pi$ and $K$ and baryons $p$,
$n$ and $\Lambda$.
The numerical estimates based on some recent parametrizations of FFs
confirm these constraints and also are in agreement with the truncated
contributions to the charge and isospin sum rules.

We also discussed the possible cancellations of Wilson line
contributions between quarks and antiquarks for valence fragmentation
functions to resolve the problem of violation of the charge sum rules
for FFs coming from the small-$z$ region.

Finally, we presented a possible modification of the charge sum rule
incorporating the so-called "orphan quark" FFs and discussed its impact
on our phenomenological tests.

\section*{Acknowledgments}
We thank A.V. Kotikov,  A. Kotlorz and S.V. Mikhailov
for stimulating discussions and interesting comments.

%\bibliography{ref}
%\bibliographystyle{apsrev}

\end{document}